\documentclass[aps,pra,twocolumn,groupedaddress,longbibliography]{revtex4-1}
\usepackage{graphicx}
\usepackage{caption}
\DeclareCaptionLabelSeparator{dot}{. }
\makeatletter
\def\justified{
	\let\\\@normalcr
	\@rightskip\z@skip \rightskip\@rightskip
	\leftskip\z@skip
	\parindent 0em\relax
	\setlength{\parfillskip}{0pt plus 1fil}}
\DeclareCaptionJustification{justified}{\justified}
\usepackage{subcaption}
\usepackage{amsmath}
\usepackage{mathrsfs} 
\usepackage{braket}
\usepackage{units}
\usepackage{ragged2e}
\usepackage[colorlinks,urlcolor=blue ,citecolor=blue ,linkcolor=blue ]{hyperref}
\usepackage{xcolor}
\usepackage{nicefrac} 
\usepackage{lipsum}
\usepackage{nameref}
\usepackage{hyperref}
\usepackage{textcomp} 
\usepackage{urwchancal}
\usepackage{xcolor}
\usepackage{float}
\definecolor{darkgreen}{rgb}{0,0.5,0}
\usepackage{multirow}
\usepackage{tabularx}
\usepackage{ulem}

\newcommand{\bs}{\boldsymbol}

\newcommand{\Gred}{\ensuremath{\Gamma_{626}} }
\newcommand{\kred}{\ensuremath{k_{626}} }
\newcommand{\Gblue}{\ensuremath{\Gamma_{421}} }
\newcommand{\kblue}{\ensuremath{k_{421}} }

\newcommand{\vcap}{\ensuremath{v_{\rm cap}}}
\newcommand{\Rcap}{\ensuremath{R_{\rm cap}}}

\newcommand{\Nsat}{\ensuremath{N_{\rm sat}}}
\newcommand{\Nfour}{N_{4{\rm s}}}

\newcommand{\dMOT}{\ensuremath{\delta_{\rm 3D}}}
\newcommand{\bMOT}{\ensuremath{b'_{\rm 3D}}}

\newcommand{\PMOT}{\ensuremath{P_{\rm 3D}}}
\newcommand{\NMOT}{N}
\newcommand{\phiMOT}{\ensuremath{\phi_{\rm 3D}}}
\newcommand{\dtwoMOT}{\ensuremath{\delta_{\rm 2D}}}
\newcommand{\btwoMOT}{\ensuremath{b'_{\rm 2D}}}
\newcommand{\PtwoMOT}{\ensuremath{P_{\rm 2D}}}
\newcommand{\phitwoMOT}{\ensuremath{\phi_{\rm 2D}}}
\newcommand{\dpush}{\ensuremath{\delta_{\rm push}}}
\newcommand{\Ppush}{\ensuremath{P_{\rm push}}}
\newcommand{\kpush}{\ensuremath{k}}
\newcommand{\Gpush}{\ensuremath{\Gamma}}

\newcommand{\Isat}{\ensuremath{I_{\rm sat}} }
\newcommand{\tload}{\ensuremath{t_{\rm load}} }

\newcommand{\ttof}{t_{\rm TOF}}

\newcommand{\Dyo}{\ensuremath{^{162}}{\rm Dy} }
\newcommand{\Dyf}{\ensuremath{^{163}}{\rm Dy} }
\newcommand{\Dyff}{\ensuremath{^{161}}{\rm Dy} }

\newcommand{\uK}{\mu{\rm K}}

\newcolumntype{Y}{>{\centering\arraybackslash}X}

\begin{document}
	
	\title{A two-dimensional magneto-optical trap of dysprosium atoms as a compact source \\ for efficient loading of a narrow-line three-dimensional magneto-optical trap}
	\author{Shuwei Jin$^{1,\dagger}$, Jianshun Gao$^{1,\dagger}$, Karthik Chandrashekara$^{1}$, Christian Gölzhäuser$^{1}$, Joschka Schöner$^{1}$,  Lauriane Chomaz$^{1,\star}$.}
	
	\affiliation{%
		$^{1}$ Physikalisches Institut, Universität Heidelberg, Im Neuenheimer Feld 226, 69120, Heidelberg, Germany.
	}	

	\date{\today}
	
\begin{abstract}
 We report on a scheme for loading dysprosium atoms  into a {narrow-line} three-dimensional magneto-optical trap (3D MOT). Our innovative approach replaces the conventional Zeeman slower with a 2D MOT operating on the broad 421-nm line to create a high-flux beam of slow atoms.  Even in the absence of a push beam, we demonstrate efficient loading of the 3D MOT{, which operates on the narrower 626-nm intercombination line}. Adding push beams working at either 421 nm {or} 626 nm, significant enhancement of the loading rate is achieved. We reach the best performance, with an enhancement factor of $3.6$, using a push beam red-detuned to the 626-nm line. With loading rates greater than $10^8$\,atoms/s achieved at a moderate oven reservoir temperature of $800\,^{\circ}$C, our method offers similar or greater performance than Zeeman-slower-based systems. Our 2D-MOT-based approach 
 constitutes a promising first step for state-of-the-art quantum gas experiments with several advantages over the Zeeman-slower-based setup and is readily adaptable to other open-shell lanthanides. 
\end{abstract}

\maketitle

\section{Introduction}

Over the last decade, ultracold gases of open-$f$-shell lanthanide atoms, such as Er and Dy, have become a platform of choice for studying novel quantum phenomena~\cite{Chomaz2022dpa}. Their electronic ground state's structures provide these atoms with remarkable properties: 
the presence of a closed outer $6s$ shell yields electronic transitions with similar properties to those of Yb or Sr. The open $4f$ shell confers on these atoms an even wider variety of transitions, a large effective spin, and a large magnetic moment, among the largest of the periodic table. In particular, the latter feature allows to explore the quantum effects of long-range and anisotropic interactions using ultracold gases of open-shell lanthanides~\cite{Chomaz2022dpa}. 

The spectrum complexity brought up by the open-shell character of the magnetic lanthanides had however refrained the scientific effort to laser-cool such species for years. In 2006, a breakthrough experiment by McClelland and Hanssen~\cite{McClelland2006lcw} demonstrated the possibility of laser-cooling Er on the {tens-of-MHz-}broad transition at 401\,nm, despite the existence of numerous decay channels. This work paved the way for many experiments bringing open-shell lanthanides to ultracold temperatures~\cite{Chomaz2022dpa,Berglund2008nlm,Youn2010dmo,Lu2010tud,Sukachev2010mot,Frisch2012nlm,Maier2014nlm,Miao2014mot,Dreon2017oca, Cojocaru2017lac, Ilzhofer2018tsf, Inoue2018mot,Phelps2020ssp, Lunden2020etc} and quantum degeneracy~\cite{Lu2011sdb,Lu2012qdd,Aikawa2012bec,Aikawa2014rfd,Davletov2020mlf}.

\begin{figure*}
    \includegraphics[width = 0.95\textwidth]{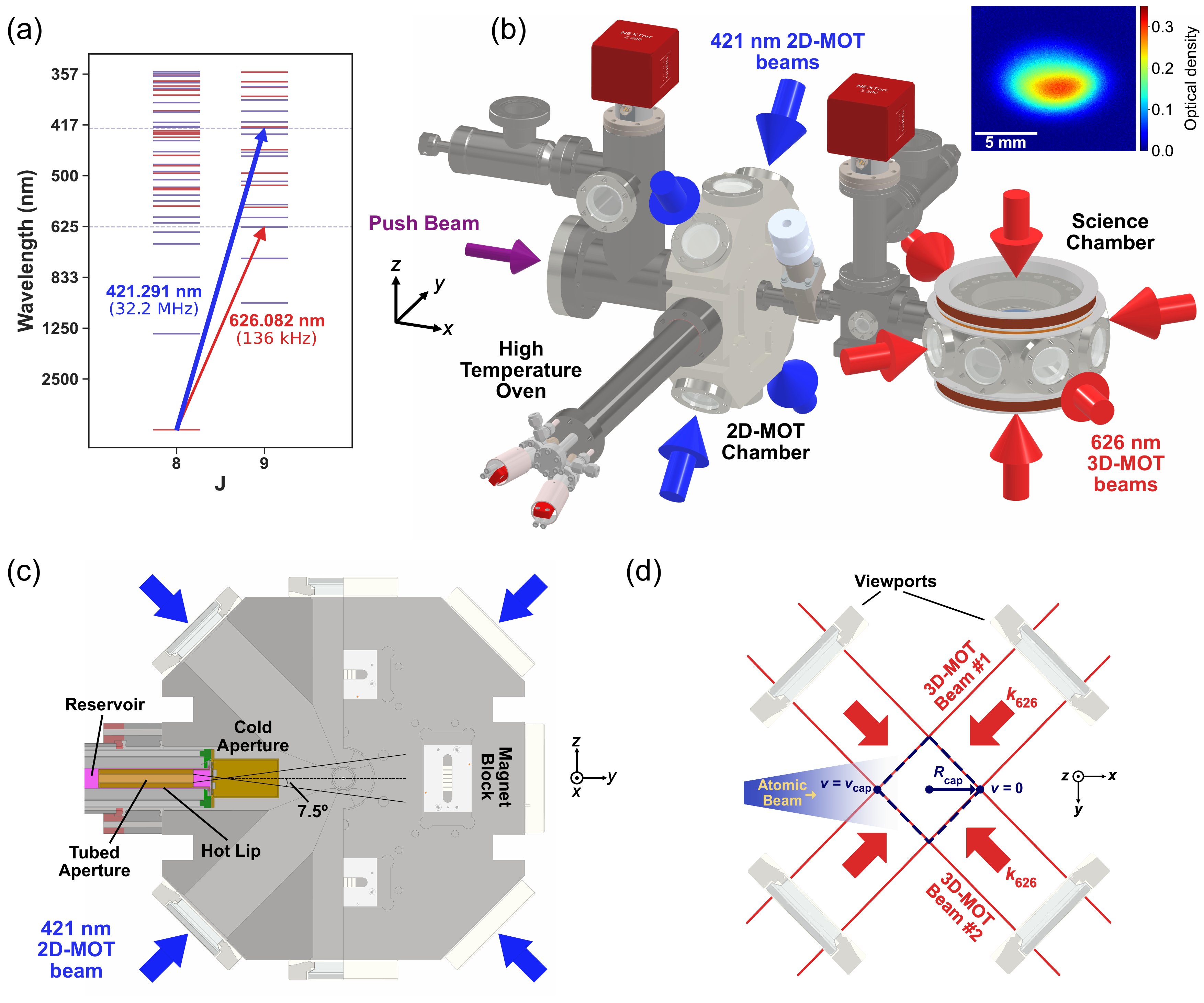}
    \caption{\label{Fig1} \textbf{Experimental Setup.} (a) Energy level diagram of Dy in wavelength $\lambda$ for the levels of total electronic angular-momentum quantum number $J=8$ and $J=9$. The state's color indicates its parity (blue, odd; red, even). The used transitions are marked by the arrows. (b) Sketch of the experimental setup showing the main vacuum components and optical paths. inset, typical absorption image of a 3D MOT. (c) Section view of the setup at the oven and 2D-MOT chamber. The aperture set at the effusion cell output is illustrated as well as an instance of magnet block. (d) Sketch of the 3D-MOT capture process within our experimental geometry. The atoms in the jet entering with $v\leq \vcap$ can be stopped within the capture region of radius $\Rcap$. This region is limited by the beam extent, and therefore the viewports' clear aperture.} 
\end{figure*}

 A particular advantage of the lanthanides is the existence of closed narrow transitions {(i.e.\, of linewidth comparable to or smaller than the recoil frequency, see e.g.~\cite{Dalibard2015ubh})}. These transitions allow for reaching ultra-low temperatures even within the Doppler cooling regime and have been exploited in laser cooling schemes. In particular, magneto-optical traps (MOTs) of open-shell lanthanides working on ultra-narrow lines have been loaded from broad-line three-dimensional (3D) MOTs~\cite{Berglund2008nlm,Youn2010dmo,Lu2010tud}, similarly to Sr and Yb~\cite{Katori1999mot,Kuwamoto1999mot}. Among all the closed narrow transitions of lanthanides, the intercombination line stands out by its 
 intermediate linewidth of a few hundreds of kHz, which allows for MOTs with capture velocities on the order of 10\,m/s and Doppler temperatures in the few $\mu$K range. Following works on Yb~\cite{Kuwamoto1999mot}, MOTs working on the intercombination line have been directly loaded from slow atomic beams, enabling a simplified cooling scheme~\cite{Chomaz2022dpa,Frisch2012nlm,Maier2014nlm,Dreon2017oca, Cojocaru2017lac}.

Another particular feature of open-shell lanthanides is their high melting point $\sim 1000\,\,^{\circ}$C. To produce the slow atomic beam required for 3D-MOT loading, all previous open-shell-lanthanide experiments use a similar scheme based on a high-temperature oven aligned with an axial Zeeman slower working on the atom's broadest cooling transition~\cite{Chomaz2022dpa,Berglund2008nlm,Youn2010dmo,Lu2010tud,Sukachev2010mot,Frisch2012nlm,Maier2014nlm,Miao2014mot,Dreon2017oca, Cojocaru2017lac, Ilzhofer2018tsf, Inoue2018mot,Phelps2020ssp, Lunden2020etc}. Besides Zeeman slowers, two-dimensional (2D) MOTs have been proven to be convenient high-flux sources of slow atoms for many of the laser-cooled atomic species {as demonstrated in the seminal work of ref.}~\cite{Dieckmann1998tdm}. This {has been applied to a wide range of species, in particular species using} an oven for a first vaporization stage, like Li~\cite{Tiecke2009hft}, Na~\cite{Lamporesi2013chf}, Sr~\cite{Nosske2017tdm,Barbiero2020sec}, Yb~\cite{Vaidya2015dbf,Saskin2019nlc} {and more recently even to species needing a refrigerated source like Hg}~\cite{guo2023Hg}. Among others, the advantages of 2D-MOT sources over Zeeman slowers are the high compactness of the setups and the absence of a direct view between the science chamber and the oven output. This is particularly beneficial for species like open-shell lanthanides, for which Zeeman slower light needs to be reflected from a mirror inside the vacuum setup in order to avoid material deposition on the viewport.   
In this case, a 2D MOT allows for better optical access around the science chamber and a simplified integration of a glass cell. Yet despite all these benefits, a 2D MOT has not been realized for open-shell lanthanides. 

In this paper, we report on an apparatus in which a 3D MOT of Dy atoms working on the intercombination line is loaded from a 2D MOT operating on the broadest cooling transition. We observe efficient loading of the 3D MOT with comparable rates $\phiMOT \gtrsim 10^8$ atoms/s and saturation atom numbers $N_{\rm sat} \approx 3\times 10^8$ as in Dy experiments based on Zeeman slowers, but at a lower oven operation temperature. The 3D MOT can be loaded even in the absence of an additional beam that pushes the atoms from the 2D to the 3D MOT (hereafter called the push beam). Using a push beam working on either the broad or the narrow transition, the loading of the 3D-MOT is significantly enhanced with a maximal enhancement factor of $3.6$ achieved using a red-detuned push beam close to the intercombination line. 

The paper is organized as follows. In Sec.\,\ref{sec:setup}, we briefly review the relevant characteristics of Dy atoms and the main components of our experimental setup, and present the 2D-MOT and 3D-MOT schemes. In Sec.\,\ref{sec:mprinciple}, the measurement scheme used for our optimization and characterization of the 2D-/3D-MOT source is described. In Sec.\ref{sec:optim2DMOT}, we report on the optimization procedure applied to our 2D-MOT source and its achievement in the absence of a push beam. In Sec.\,\ref{sec:push}, we further introduce a push beam and describe the observed enhancement of the 3D-MOT loading, comparing broad- and narrow-line push configurations. 
Finally, in Sec.\,\ref{sec:3DMOT}, the optimal 3D-MOT parameters for its loading in the absence and presence of push beams are investigated, compared, and comprehended in relation to the velocity features of the 2D-MOT atomic beam and of the 3D-MOT capture process.

\section{Experimental setup and 2D/3D-MOT scheme\label{sec:setup}}

{The Dy electronic levels and transitions of interest for this work are depicted in Fig.\,\ref{Fig1} (a) and connect its ground state of total angular-momentum quantum number $J=8$ to excited states with $J'=9$~\cite{Martin1978ael}. The broad transition of wavelength $\lambda_{\rm 421}=2\pi/\kblue=421.291\,$nm has a linewidth of $\Gamma_{\rm 421}=2\pi\times 32.2\,$MHz and a saturation intensity of $\Isat^{421}=564\,\textrm{W/m}^2$. The narrower intercombination line of wavelength $\lambda_{\rm 626}=2\pi/\kred=626.082\,$nm has a linewidth of $\Gamma_{\rm 626}=2\pi\times 136\,$kHz and a saturation intensity of $\Isat^{626}=720\,\textrm{mW/m}^2$. The ground state, 421-nm and 626-nm excited states have similar $g$ factors of $g_J=1.24$, $g_{J'}^{(421)}=1.22$, and $g_{J'}^{(626)}=1.29$ respectively.}

The 421-nm and 626-nm laser lights used to address these transitions are generated by two commercial frequency-doubled amplified diode lasers (DLC TA-SHG PRO) from TOPTICA Photonics AG. The lasers are operated with 800mW and 1.6W output power respectively. 
Both lasers are frequency-locked to a commercial ultra-low expansion cavity from Stable Laser Systems using the offset-sideband Pound-Drever-Hall scheme~\cite{Black2001ait}. The cavity has a 1.46\,GHz free spectral range and a finesse determined by means of cavity-ring-down measurements to be 21170(60) at 842\,nm and 20760(40) at 626\,nm, corresponding respectively to the wavelengths used to lock the 421-nm and 626-nm lasers.

The vacuum apparatus for our experiment is depicted in Fig.\,\ref{Fig1} (b). It consists of one high-temperature dual-filament effusion cell (DFC-40-25-WK-2B) from CreaTec Fischer \& Co. GmbH, and of two chambers -- the 2D-MOT chamber and the science chamber -- connected via a differential pumping section. Most vacuum components have been produced by SAES Rial Vacuum out of stainless steel (grade 316L or 316LN) or titanium. Pressures on the order of $10^{-9}$\,mbar and $10^{-11}$\,mbar are achieved in the 2D-MOT and science chambers respectively. 

Solid Dy pieces are placed in the reservoir region of the effusion cell and are vaporized by heating this region up to $800\,^{\circ}$C. {The relatively low temperature of the reservoir region was chosen to spare material and allow for a long lifetime of the source. Figure\,\ref{Fig1} (c) details the setup design from the effusion-cell output to the 2D-MOT chamber. The oven is inserted into the 2D-MOT chamber, transversely to the 2D-MOT axis ($x$-axis). The distance between the oven's last aperture and the center of the 2D-MOT chamber (40.8mm) is minimized with the aim of maximizing the incoming atomic flux in the 2D-MOT. 
A Dy atomic vapor jet is formed by a custom-designed set of apertures located in the hot-lip region of the effusion cell, which is heated up to $1100\,^{\circ}$C. A last cold aperture, connected to the oven water-cooling system, enables to filter out the part of the atomic jet exiting the oven at angles larger than $7.5\,^{\circ}$.}  

The atoms exiting the oven into the 2D-MOT chamber experience the forces induced by the cooling beams. The cooling beams are made up of a single 421-nm laser beam propagating through the chamber in a bow-tie $\sigma_+\sigma_-$ retro-reflected configuration. The bow-tie plane ($yz$-plane), together with the magnetic-field configuration (see below), defines the 2D-MOT axis as the orthogonal $x$ axis (see Fig.\,\ref{Fig1} (b,c)).  
The laser beam has a power of $430\,$mW and waist of $16\,$mm~\footnote{{In the main-text measurement, we use the maximal power available while the 2D-MOT power dependence is reported in App.~\ref{app:2DPow}}.}. The laser beam  path does not include any active optical components in order to ensure a high effective power on the atoms. Therefore, the beam frequency can only be altered through the laser locking point, and its power via mechanical means. 

Eight stacks of permanent magnets are placed symmetrically on both sides of the 2D-MOT chamber, as partially illustrated in Fig.\,\ref{Fig1} (c), to provide the 2D-MOT magnetic field. The magnetic field is zero-valued along the 2D-MOT axis and oriented along the cooling beams on their propagation axes. It has a roughly uniform gradient in the $yz$ plane over the chamber's central region whose magnitude, $\btwoMOT$, can be adjusted by changing the number of magnets. Hereafter, the values of $\btwoMOT$ correspond to the theoretical expectations for a perfect arrangement of magnet stacks and no other magnetic source. Increasing the number of magnets by one in each of the 8 blocks increases the gradient by approximately $4.4$\,G/cm. A gradient of up to $44.4$\,G/cm can be generated with our current magnet-holder design. To optimize the performance of the 2D MOT, we adjust the divergence of the cooling beam, its detuning from resonance $\dtwoMOT$, as well as the number and positions of the magnets. We note that due to the permanent magnets implemented, adjusting the 2D-MOT magnetic field involves an important manual aspect.  

The atoms trapped in the 2D MOT can travel along the $x$ direction and reach the center of the science chamber. The distance between the two chamber centers along $x$ was minimized during our design process. It equals $347$\,mm and includes a $55.7$\,mm-long differential pumping section, which is inserted into the 2D-MOT chamber. {The differential pumping section is of conical shape and has an aperture diameter of 2$\,$mm on one end and 5.9$\,$mm on the other.} At the center of the science chamber, a 3D MOT is formed using three orthogonal retro-reflected 626\,nm laser beams and a pair of magnetic coils. The coils are aligned along the $z$-axis and connected in anti-Helmholtz configuration to provide a magnetic gradient $\bMOT$ up to 4\,G/cm in the current configuration. Hereafter, the values of $\bMOT$ correspond to the gradient in the $xy$ plane extracted from numerical calculations using our coil geometry. 
 In the present setup an additional pair of magnetic coils, aligned along the $z$ axis and connected in Helmholtz configuration can be used to apply a tunable offset magnetic field along the $z$ direction, aligned with gravity. 

An important constraint related to the narrow-linewidth transition used for our 3D-MOT scheme is the low capture velocity of the 3D MOT, $\vcap$. {The principle of the capturing process is illustrated in Fig.\,\ref{Fig1} (d): atoms of the 2D-MOT jet with an initial axial velocity $v_x<\vcap$ have to be stopped by the 3D-MOT radiation pressure within the 3D-MOT capture region of radius $\Rcap$.} An upper bound on $\vcap$ can be estimated in the limit of an infinitely saturated transition, where the atoms scatter photons at a rate $\Gred/2$ independent of the light detuning. In this approximation, a constant radiation pressure force is exerted onto the atoms over the 3D-MOT capture region, yielding $\vcap \leq \sqrt{2\Rcap\hbar\Gred k_{626,x}/{m}}$, with $k_{626,x}$ the recoil momentum transferred by one photon along the $x$ axis ($\hbar$ is the reduced Planck's constant), see e.g.\,Ref~\cite{Dalibard2015ubh}. With our geometry [see Fig.\,\ref{Fig1} (d)], $k_{626,x} = \kred/\sqrt{2}$ and even with MOT-beam sizes equal to the viewports' clear aperture (yielding $\Rcap  = 35/\sqrt{2}$\,mm), we estimate the maximum capture velocity of our 626-nm 3D MOT to be $\vcap \lesssim 11\,$m/s.  To favor the loading of atoms into the 3D MOT, we thus use relatively large {3D-MOT beam waists of $w_{\rm 3D}= 12\,$mm with a power of $\PMOT \approx 85$\,mW per beam, so as to nearly fulfill the above estimates of the capture radius and velocity. 
Note that the atoms fall under gravity when traveling from the 2D-MOT to the 3D-MOT chamber, which might compromise their capture. With a horizontal velocity of 11\,m/s, the falling distance is 4.9\,mm and is smaller than $\Rcap$. Furthermore, if the atoms emerge from the 2D-MOT with an 11\,m/s velocity oriented 15\,mrad upward, the fall is suppressed.  
The relevant parameters for the optimization of the 3D-MOT loading are} the detuning of the 3D-MOT beams, $\dMOT$, and the magnetic-field gradient, $\bMOT$.

 In the following, we focus on the isotope $^{164}$Dy {(atom mass $m=164\,$amu with amu the atomic mass unit)}, which has the highest natural abundance. To gain first insights into the  expected performance of our 2D-MOT-based source and the relevant parameter ranges, we perform Monte Carlo simulations of the 2D- and 3D-MOT capture processes along the lines of ref.~\cite{Barbiero2020sec}, see Appendix \ref{app:sim2D} and ref.~\cite{Jianshun2022thesis} for details. Using an oven temperature of $T=1000\,^{\circ}$C, our simulations indicate that a maximal 2D-MOT flux is achieved for $\btwoMOT= 31\,$G/cm and $\dtwoMOT= -2.1\,\Gblue$ and is estimated to $\phitwoMOT \approx 3 \times 10^{10}$ atoms/s for $^{164}$Dy. The 3D-MOT loading process can also be included in the simulation. However, due to the large fraction of unloaded trajectories over the full process and our limited number of simulated trajectories (see Appendix \ref{app:sim2D}), the extracted 3D-MOT loading rates show large fluctuations. In the simulations, loading into the 3D-MOT is detected when adding a push beam and 3D-MOT loading rates on the order of $9 \times 10^8$ atoms/s were extracted for $^{164}$Dy. Based on these simulation results and their parameters, we started our experimental search. In the experiment, we observed a 3D-MOT loading even without an additional push beam. This experimental observation serves as the starting point for our optimization process.

\section{Measurement protocol} \label{sec:mprinciple} 

In the following, we characterize the performance of our 2D-MOT-based atom source through the achieved 3D-MOT loading rates and atom numbers. Our experimental procedure is as follows: In a first step we switch on the 2D-MOT, optional push, and 3D-MOT beams, the offset-field and gradient coils, with fixed parameter values for a time $\tload$. We then switch off the 2D-MOT and push beams, and hold the cloud for 70\,ms without changing any other parameter values except the offset field~\footnote{The experiment is performed in the presence of a typical remnant bias field of 0.4\,G mostly directed along $z$. Applying a $z$-bias field was found to have only a marginal effect on the 3D-MOT loading performances when varied over a few-G range and will not be further discussed here. It is typically set to 0.4\,G at the loading stage.}. Finally, we switch off the 3D-MOT light and gradient field, let the cloud fall and expand for a short time of flight (TOF) $\ttof$ (typically $\ttof=5\,$ms), and take an absorption image using horizontally linearly polarized 421-nm light propagating along the $y$ axis. 
The absorption signal is recorded on a CMOS camera (Hamamatsu Orca Spark) via an imaging lens providing a magnification of $0.438$. {The imaging pulses last for $25\,\mu$s}. The imaging beam is operated on resonance, with an intensity below $0.2\,\rm{mW}/\rm{cm}^2$ and a waist of about $10\,$mm.  
An exemplary image is shown in the inset of Fig.\,\ref{Fig1} (b).  

For the present characterization, we do not compress the gas by decreasing the light detuning and intensity after the 3D MOT loading stage. Therefore the cloud has a relatively high temperature of about $500\,\uK$ while the remnant field is estimated to be around $0.4\,$G. At these temperature and magnetic-field values, the atomic population is expected to be depolarized, and all Zeeman substates occupied.
Assuming 
an equal population of all substates, the light-scattering cross-section, $\sigma$, is identical for any light polarization and is given by renormalizing the bare cross-section $\sigma_0=3\lambda_{421}^2/2\pi$ by the average of the {squares of the} Clebsch-Gordan coefficients for our $J=8 \rightarrow J'=9$ dipole transition, over the initial Zeeman substates. This yields $\sigma=0.3725\sigma_0$.  
We observe that the absorption signal does not depend on 
the imaging light polarization, which experimentally supports the use of $\sigma=0.3725\sigma_0$. 

By integrating over a region of interest in the absorption images, we extract the atom number $\NMOT$ in the 3D MOT at the end of the sequence. {The statistical uncertainty on $\NMOT$ is typically of 5\% and we have estimated the systematic uncertainty to be of about 4\% with dominant contributions from the imaging magnification and unequal populations of Zeeman substates at finite temperature.} To optimize our setup, we mostly rely on a simple scheme that consists in measuring $\NMOT$ for a characteristic $\tload =4\,$s, hereafter referred to as $\Nfour$. To further characterize our system, we record loading curves of $\NMOT$ versus $\tload$, which we fit to an exponential growth function, $\NMOT(t) =\Nsat (1-e^{-t/\tau})$, to extract the 3D-MOT loading rate, $\phiMOT = \Nsat/\tau$, and the saturation atom number $\Nsat$. 

We note that the resonance frequencies for the 421-nm and 626-nm lights are extracted from the 3D-MOT images.  The 421-nm resonance frequency is directly extracted from the maximum in the absorption signal when scanning the frequency of the imaging light. The resonance frequency for the 626-nm light is determined by shining an additional 626-nm beam onto the atoms during the first milliseconds of their time of flight and monitoring the atom-number depletion versus the beam frequency via absorption imaging after time of flight.

\section{Optimization of the 2D MOT, without push beam} \label{sec:optim2DMOT} 

In our setup, the optimization of the 2D-MOT parameters has been performed, following the first observation of a 3D-MOT loading, in the absence of a push beam. 
Here we report on the protocol followed in the optimization process and the performance achieved in this configuration. We start the optimization process by setting the 2D-MOT parameters close to the simulated optimum (see Sec.\,\ref{sec:setup}), where a 3D-MOT loading is detected ($\btwoMOT = 35.4\,$G/cm, $\dtwoMOT=-2\Gblue$). We scan the 3D-MOT gradient and detuning to maximize the 3D-MOT loading, yielding $\bMOT = 0.9$\,G/cm and $\dMOT = -55.8(3)\Gred$. 
Using this 3D-MOT setting, we go on by changing the 2D-MOT parameters as follows: choosing a given number of magnets per stack, we start by setting the magnet stacks at their design positions.  
In this configuration, we optimize the divergence of the 2D-MOT cooling beam~\footnote{We note that the beam divergence is particularly sensitive because the power balancing between counter-propagating beams in the 2D MOT is relevant not only at the center but all along its axis and in particular close to the differential pumping stage.} and its detuning $\dtwoMOT$ by maximizing $\Nfour$. We then adjust the position of each magnet stack and iterate on the beam parameters.    

\begin{figure}
    \includegraphics[width = 0.45\textwidth]{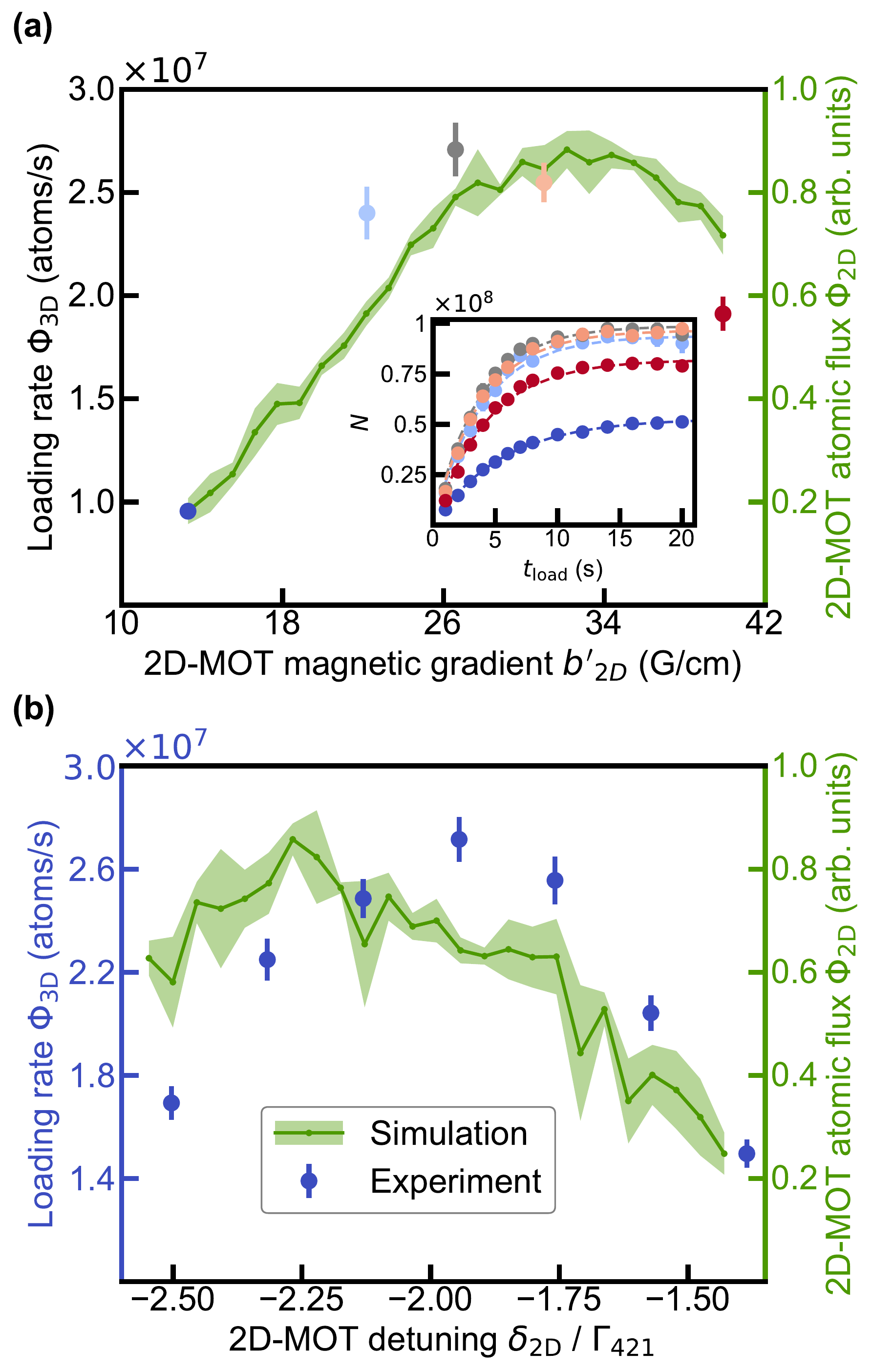}
    \caption{\label{Fig2} \textbf{2D-MOT optimization without push beam}.(a) Experimental 3D-MOT loading rate $\phiMOT$ (circles, left axis) and simulated 2D-MOT flux $\phitwoMOT$ (green line and dots, right axis) as a function of the 2D-MOT magnetic gradient $\btwoMOT$. The 3D-MOT parameters were fixed to $\bMOT = 0.9$\,G/cm, $\dMOT = -55.8(3)\Gred$ and the 2D-MOT configurations (including $\dtwoMOT$) were individually optimized, see text. The inset shows the experimental 3D-MOT loading curves and exponential fits from which the rates are extracted, and the errorbars show the standard deviation of three experimental runs. (b) Experimental $\phiMOT$ (blue circles) and simulated $\phitwoMOT$ (green line and dots) as a function of the 2D-MOT  detuning $\dtwoMOT$ at the experimental optimal gradient, $\btwoMOT = 26.7\,$G/cm. The 3D-MOT parameters were fixed to $\bMOT = 0.42$\,G/cm, $\dMOT = -42.6(3)\Gred$. In (a) and (b), the errorbars are the $63\%$ confidence interval from the fit. The shaded area shows the standard deviation of three simulation runs.}
\end{figure}

For each number of magnets per stack implemented, we identify the best values of the 2D-MOT parameters and we finally record the full 3D-MOT loading curve in the optimized configuration. Examples of such loading curves and their fits (see Sec.\,\ref{sec:mprinciple}) are shown in Fig.\,\ref{Fig2} (a) inset. 
Figure\,\ref{Fig2} (a) displays the dependence of the fitted loading rate $\phiMOT$ with the magnetic gradient $\btwoMOT$. A maximum loading rate of  $\phiMOT = 2.7(2)\times 10^7$ atoms/s and a saturation atom number of $\Nsat = 9.9(2)\times 10^7$ is found for the configuration with 6 magnets per stack ($\btwoMOT \approx 26.7\,$G/cm) and $\dtwoMOT=-1.95(1)\Gblue$.
We compare these experimental observations to the expectations from our Monte Carlo simulations, see Sec.\,\ref{sec:setup}, and Appendix \ref{app:sim2D} for details. So as to obtain reliable results despite limited sampling, we do not include the 3D-MOT loading step and simulate only the 2D-MOT loading process.  
Figure\,\ref{Fig2} (a) shows the simulated 2D-MOT flux as a function of $\btwoMOT $ with an oven at $T=800\,^{\circ}$C and at the value of $\dtwoMOT$ for which this rate is maximal, like in the experimental procedure. 
We observe that the optimal $\btwoMOT $ and $\dtwoMOT$ are comparable yet respectively slightly smaller and larger in magnitude for the experimental $\phiMOT$ compared to the simulated $\phitwoMOT$ (optimum at $\btwoMOT = 32\,$G/cm and $\dtwoMOT=-2.1\Gblue$). Both in experiment and theory, we find that the optimum value of $|\dtwoMOT|$ slightly varies over the explored $\btwoMOT$-range, by about $0.2\,\Gblue$ and $0.7\,\Gblue$ respectively. 
The similarity between the experiment and simulation results is remarkable given the approximations made in the simulations, see Appendix \ref{app:sim2D}. The simulations do not necessarily provide the quantitative optima for different parameters, but are a suitable tool to identify the relevant range of parameters, see Sec.\,\ref{sec:setup}. 

In the optimized magnetic configuration identified above, we further investigate the influence of the 2D-MOT detuning and record the full 3D-MOT loading curve for different values of $\dtwoMOT$. The extracted loading rates are shown in Fig.\,\ref{Fig2} (b). Over the investigated $\dtwoMOT$ range ($1\Gblue$), the variations of $\phiMOT$ are rather symmetric around its maximum, and its value changes by less than $50\%$. {In Fig.\,\ref{Fig2} (b), we also show the simulated 2D-MOT flux at $\btwoMOT = 26.7\,$G/cm. The variations of the simulated $\phitwoMOT$ match those of the experimental $\phiMOT$ well.}
Experimentally, the maximum is found at $\dtwoMOT=-1.95(1)\Gblue$ with a loading rate of  $\phiMOT = 2.7(1)\times 10^7$ atoms/s and a saturation atom number of $\Nsat = 12.2(1)\times 10^7$. We note that different 3D-MOT parameters were used compared to Fig.\,\ref{Fig2} (a), which explains the slightly different loading performances (see also Sec.\ref{sec:3DMOT}). 

\section{Push-beam enhancement}\label{sec:push}

\begin{figure*}
    \includegraphics[width = \textwidth]{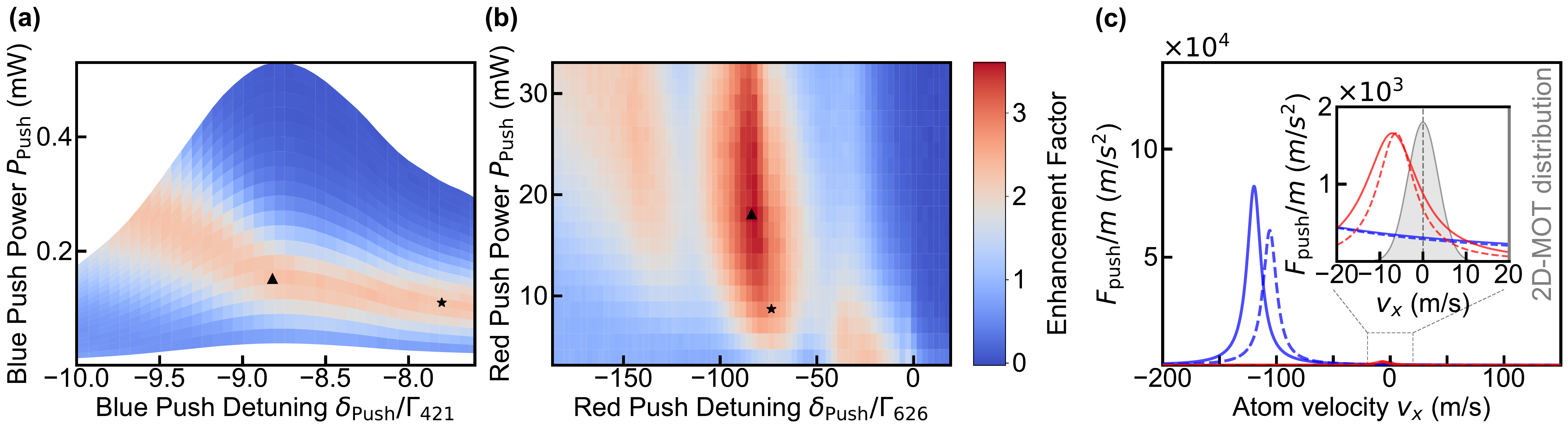}
    \caption{\label{Fig3} \textbf{Push-beam Effect}. Enhancement factor in the atom number in 3D-MOT at $\tload=4s$, $\Nfour$, obtained by the addition of a 421-nm (a) and 626-nm (b) push beam, as a function of the push-beam power $\Ppush$ and detuning $\dpush$. The 3D-MOT parameters were set to $\bMOT  = 0.42$\,G/cm, $\dMOT = -42.6(3)\Gred$. 
    The reference without push beam is $\Nfour=7.5(3)\times 10^7$.
    $\Ppush$ is tuned through the amplitude of the driving signal of an acousto-optic modulator. The $\Ppush$ values are extracted via a rescaling of the measured (for (a), frequency-dependent) power in front of the entrance viewport at fixed diffraction efficiency. (c) Velocity-dependent force \eqref{Equ:push_force} exerted by the 421-nm (blue lines) and 626-nm (red lines) push beams for the characteristic parameters highlighted by triangle (solid lines) and star (dashed lines) symbols in (a) and (b), respectively. The forces are calculated assuming no magnetic field. The inset shows a zoom-in in the relevant low-velocity region. The grey-shaded area illustrates a Gaussian approximation of the expected velocity distribution of the 2D-MOT atomic beam.     
    }
\end{figure*}


To further increase the loading rate of our 3D MOT, we implement a push-beam scheme, as typically done in 2D-MOT setups, see e.g.\,refs.~\cite{Dieckmann1998tdm,Tiecke2009hft,Lamporesi2013chf,Nosske2017tdm, Barbiero2020sec,Dorscher2013coq,Vaidya2015dbf,Saskin2019nlc,guo2023Hg,Kwon:2023}: we additionally shine a beam propagating through the apparatus along $+x$, see Fig.\ref{Fig1} (b). This beam has a frequency close to one of the transitions of Dy and, via radiation pressure, pushes the atoms from the 2D-MOT to the science chamber in a velocity-selective way.  
In the case of Dy, similarly to Yb~\cite{Dorscher2013coq,Vaidya2015dbf,Saskin2019nlc}, two convenient choices are possible: the push beam can be near-resonant either with the 626-nm intercombination line or with the broad 421-nm one. To determine which choice is more beneficial, we implemented both sequentially and compared their experimental achievements. We note that, in both cases, a beam-walking optimization yields a configuration in which the push beam goes through the differential pumping stage and is detected on the opposite side of the science chamber. The push-beam waist is $w_{\rm push}\approx 0.8$\,mm, and, in order to control the effect of the push beam, we vary its power $\Ppush$, and detuning $\dpush$. In either case (421-nm or 626-nm), an optimization on the push-beam parameters  clearly improves the 3D-MOT loading. 

\begin{figure*}
    \includegraphics[width = \textwidth]{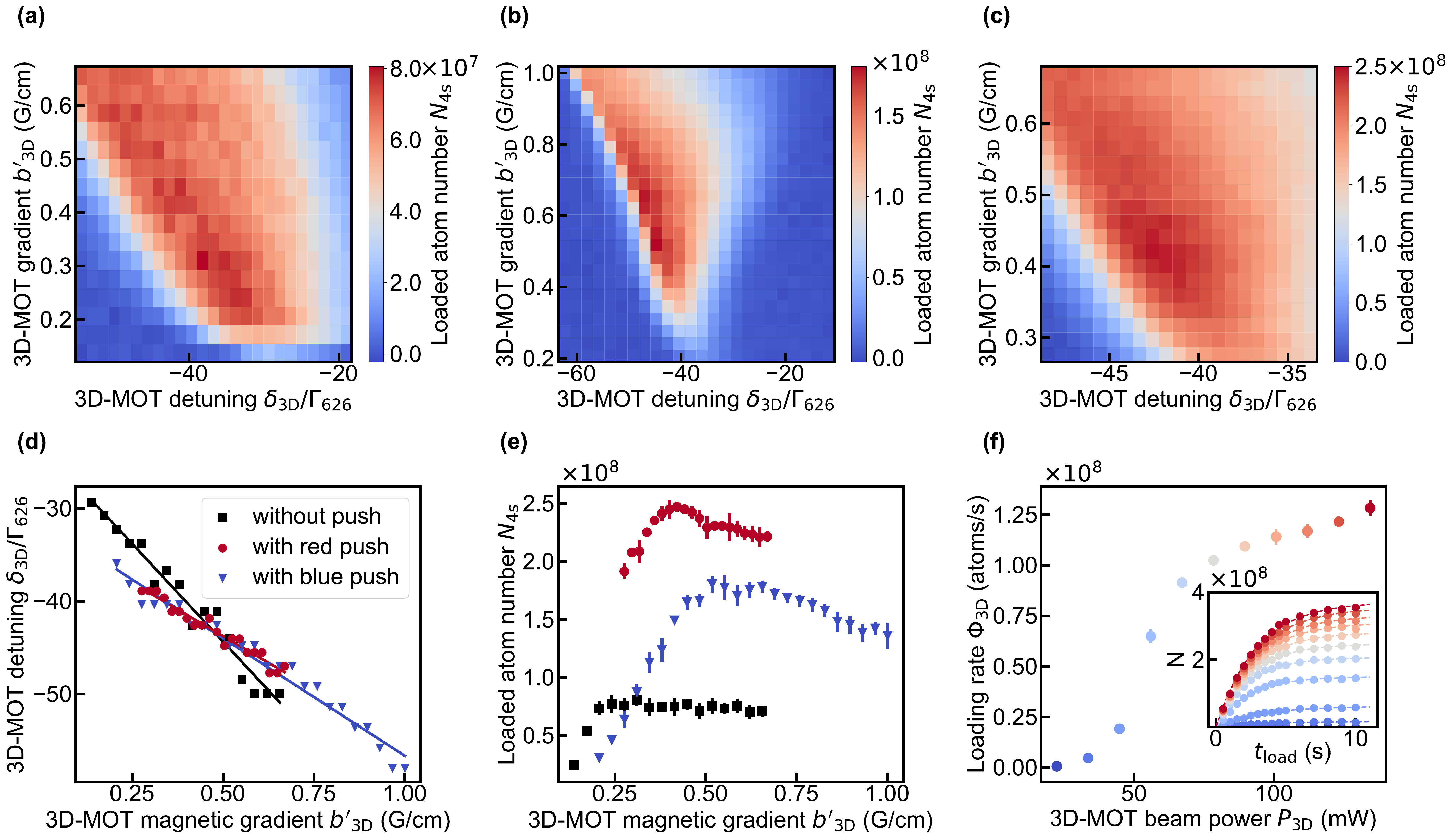}
    \caption{\label{Fig4} \textbf{3D-MOT with and without push beam}. (a-c) Experimental 3D-MOT atom number, $\Nfour$, as a function of the 3D-MOT detuning $\dMOT$ and gradient $\bMOT$, in (a) the absence, or the presence of a (b) 421-nm and (c) 626-nm push beam.  The other parameters are set to the optimal values identified in Figs.\,\ref{Fig2} and \ref{Fig3}. In particular, we set (b) $\dpush=-8.3\Gblue$, $\Ppush=0.26$\,mW and (c) $\dpush=-82.3(3)\Gred$, $\Ppush=12$\,mW.   (d) 3D-MOT detuning, $\dMOT^*$, at which the maximal $\Nfour$ is achieved at fixed $\bMOT$. The values are extracted from the scans of (a, black square), (b, blue triangle), (c, red circle) and shown as a function of $\bMOT$. (e) Optimal values of $\Nfour$, $N^*$, as a function of $\bMOT$. Same code as (d). (f) 3D-MOT loading rate as a function of the 3D-MOT beam power, $\PMOT$, with a 626-nm push beam of $\dpush=-82.3(3)\Gred$, $\Ppush=18$mW, and $\bMOT  = 0.42$\,G/cm, $\dMOT = -42.6(3)\Gred$. The inset shows the corresponding full 3D-MOT loading curve (circles) and their exponential fits (lines). The different colors correspond to the different $\PMOT$ values. In (d-f), the error bars show the standard deviation of three experimental repetitions. }
\end{figure*}

We perform a systematic experimental study of the 3D-MOT atom number $\Nfour$ while varying the push-beam parameters in a range where improved 3D-MOT loading is found. Figures\,\ref{Fig3} (a) and (b) show the enhancement factor {defined by the ratio of $\Nfour$ obtained in} a configuration with and without a push beam {operating at [Fig.\,\ref{Fig3} (a)] 421 nm and [Fig.\,\ref{Fig3} (b)] 626 nm}. 
Here we use $\bMOT  = 0.42$\,G/cm, $\dMOT = -42.6(3)\Gred$, and the reference without push beam is $\Nfour=7.5(3)\times 10^7$ [see also Fig.\,\ref{Fig4} (a),(b),(c)]. 
In both Fig.\,\ref{Fig3} (a) and (b), the enhancement factor at fixed $\Ppush$ shows a maximum for varying $\dpush$. We now describe the power dependence of this maximum. 
In either case (421-nm and 626-nm), the $\dpush$ value at which the maximum is found increases in absolute value for increasing $\Ppush$. Yet, the power dependence shows different features in the two cases. In particular, the value of the maximum enhancement factor itself shows distinct variations with $\Ppush$. In the case of the 421-nm push beam, this maximum enhancement factor is nearly independent of $\Ppush$. In contrast, in the 626-nm case, it shows an overall optimum in power, corresponding to $\Ppush = 18$\,mW. Overall, the push beam on the narrow transition is found to outperform the one on the broad transition and yields the maximal enhancement factor observed of $3.6(2)$.  

These experimental findings can be comprehended from a description of the push-beam effect, which is rooted in the radiation pressure force,
\begin{eqnarray}
    \label{Equ:push_force}
 \bs{F}_\text{push}({v}_x) &=& \hbar \kpush \bs{e}_x \frac{\Gpush}{2} \frac{s_0}{(1+s_0)}\frac{1}{1 + 4\frac{\left({\dpush} - {\kpush}v_x\right)^2}{(1+s_0)\Gpush^2}}.
\end{eqnarray}
Here  $\kpush$ is the push-beam wavenumber, $s_0=2\Ppush/(\pi w_{\rm push}^2I_{\rm sat})$ is the saturation parameter, $\Gpush$ and $I_{\rm sat}$ are the associated transition's linewidth and saturation intensity. The force~\eqref{Equ:push_force} is directed along the push-beam propagation direction of unit vector $\bs{e}_x$, and its magnitude depends on the atom's velocity along $x$, $v_x$, through the Doppler effect. More precisely it is a Lorentzian function of $v_x$, of center $\dpush/\kpush$, of width $\sqrt{1+s_0}\,\Gpush/\kpush$, and of amplitude on resonance $\hbar \kpush\frac{\Gpush}{2} \frac{s_0}{(1+s_0)}$. The effects of the push-beam parameters are as follows: Varying $\dpush$ changes the velocity class with which the force is resonant. Changing $\Ppush$ alters $s_0$ and has a twofold effect: (i) it scales the resonant amplitude of the force up by the factor $\frac{s_0}{1+s_0}$ and (ii) it broadens the range of velocities addressed by the force by $\sqrt{1+s_0}$. The effects (i) and (ii) dominate at low and high saturation respectively. Note that Eq.\,\eqref{Equ:push_force} assumes no effect of the magnetic field, which theoretically cancels along the propagation axis of the push beam. 

The drastically different linewidths of the 421-nm and 626-nm transition result in a different operation of the corresponding push beams. The 421-nm push beam may generate much larger forces than the 626-nm beam. In particular, a force magnitude corresponding to the saturated 626-nm case is generated with a 421-nm beam of saturation parameter as low as $s_0 =0.0021$. Furthermore, the bare velocity widths of the force $\Gpush/\kpush$ are also different, equal to $0.085$\,m/s and $13.6\,$m/s for the 626-nm and 421-nm transition respectively. This is to be compared with the expected spread of velocity distribution of the 2D MOT of about $4\,$m/s defined by its half width at half maximum.

These different linewidths, together with the powers used in experiment~\footnote{For the 421-nm beam, the range of power extends from $\Ppush \approx 20$ to $\approx 400\, \mu$W. Lower powers were not used due to the difficulty of obtaining a stable power value in this range.}, result in broadly different force profiles in the optimal push configurations. Figure\,\ref{Fig3} (c) sketches such force profiles as expected from Eq.\,\eqref{Equ:push_force}. In both the 421-nm and 626-nm cases, improved loading conditions are achieved for red detunings, $\dpush<0$, which corresponds to a resonant pushing of atoms traveling against the push beam propagation ($v_x<0$). Yet, the resonant velocity classes are vastly different with $v_x\sim -10$\,m/s ($v_x\sim -100$\,m/s) for the 626-nm (421-nm) case. Therefore, in the 421-nm case, mostly the detuned "tail" of the radiative force is involved in pushing the atoms. 
On the contrary, the much weaker force generated by the 626-nm light is used on and close to its resonance. 

The 421-nm and 626-nm push beams also operate at considerably different saturation parameters of $s_0\leq 0.5$ and $3\times 10^3<s_0<60\times 10^3$, respectively.  Following the discussion on the effects (i)-(ii), a change in power thus affects the force profiles of the 421-nm and 626-nm lights differently. As illustrated in Fig.\,\ref{Fig3} (c) through two typical situations of different $\Ppush$, an increase in power for the 421-nm beam mostly yields a scaling up of the resonant force magnitude.  In contrast, the 626-nm push operates at roughly constant (saturated) resonant force magnitude and the change in power mainly yields a broadening of the resonance.  

Relevant to the push-beam effect on our 2D-MOT are the force profiles in the small $|v_x|$ range encompassing the 2D-MOT velocity distribution, see Fig.\,\ref{Fig3} (c) inset. The power dependences of the force in this range explain the behaviors of the maximal enhancement observed in Figs.\,\ref{Fig3} (a) and \ref{Fig3}(b). The overall shift to larger $|\dpush|$ values for increasing $\Ppush$ is justified as follows: increasing $\Ppush$ at fixed $\dpush$ yields a detrimental effect of pushing the atoms with positive $v_x$ too much (via power scaling or broadening of the force) such that the final $v_x$ may exceed the 3D-MOT capture velocity $\vcap$. Instead, shifting $\dpush$ to larger negative values prevents this effect and additionally yields the benefit of pushing atoms with larger negative $v_x$ back towards the 3D MOT. 

This shift has however different impacts in the 421-nm and 626-nm cases due to their different operation regimes. 
In the 421-nm case, the small $|v_x|$ range corresponds to a far-detuned regime in which the amplitude scaling is well compensated by a shift of the resonance. Therefore, when increasing $\Ppush$, $\dpush$ is adjusted such that the force profile in this $|v_x|$ range is kept basically unchanged. The push-beam effect is thus nearly power-independent and so is the enhancement factor.
In the 626-nm case, instead, the force profile in the optimal-enhancement condition changes with $\Ppush$ in the small $|v_x|$ range, affecting the push efficiency. At low power, the range of velocity classes addressed by the force is small compared to the velocity distribution itself, yielding a low push efficiency. As described above, increasing $\Ppush$, the 626-nm force broadens with a saturated resonant amplitude. 
Therefore if one tries to keep a constant push effect (i.e.\,force profile) in the small negative $v_x$ range while increasing $\Ppush$, the saturation and power broadening effects imply an increasing detrimental push effect on the positive $v_x$ range. Hence, at too large power, either the push effect on $v_x>0$ or on $v_x<0$ is not optimal, and an optimum efficiency is expected at an intermediate power.  At the optimum $ \Ppush = 18$\,mW, the 626-nm push force has an expected velocity width of about $7$\,m/s, comparable to the expected width of the 2D-MOT distribution.
 The fact that the 626-nm force profile can be adapted to strongly push the $v_x\lesssim 0$ class with a reduced impact on the $v_x\sim 10\,$m/s one may explain the observed better performance of the 626-nm push beam. 
{We note that a foreseeable way to further improve the loading of the 3D MOT, instead of relying on the simple power broadening of the push beam, is to introduce sidebands in the push-beam frequency, see e.g.\,Refs~\cite{Vaidya2015dbf,Saskin2019nlc,Maier2014nlm,Dreon2017oca}. This approach could allow tailoring the velocity-dependent lineshape of the force to effectively address different velocity classes.}

For the case of the 626-nm push beam, we additionally observe a rather unexpected behavior in Fig.\,\ref{Fig3} (b): $\Nfour$ shows multiple local maxima when varying $\dpush$ at fixed $\Ppush$. 
This effect may be explained by the possible presence of a remnant magnetic field along the push-beam propagation axis. Such a magnetic field makes the vectorial nature of the push-beam transition ($J=8 \rightarrow J'=9$) become relevant and modifies the simple picture of Eq.\,\eqref{Equ:push_force} and Fig.\,\ref{Fig3} (c). In particular, this yields different resonant conditions for the push-beam light components driving the $\sigma_+$, $\pi$, and $\sigma_-$ transitions respectively. Therefore, the total force profile, given by the sum of these three contributions, would then present three distinct peaks with different resonant velocities, and amplitudes and widths set by the light-polarization composition. The relevance of this effect is supported by an observed change in the relative strengths of the maxima in $\Nfour$ when changing the push-beam polarization. We note that the best performance is found with a push beam of horizontal linear polarization. 


\section{3D-MOT parameters and capture with and without push beam}\label{sec:3DMOT}

In a final study, we investigate the optimal 3D-MOT settings and their variations without, with a 421-nm, and with a 626-nm push beam. We set the push-beam parameters to the values providing the optimal enhancement factors in Fig.\,\ref{Fig3}. 
Figures \ref{Fig4} (a), (b), and (c) show the atom number $\Nfour$, while scanning the 3D-MOT detuning $\dMOT$ and magnetic gradient $\bMOT$ in these three push-beam configurations. The three plots differ in their overall magnitude but show qualitatively similar variations with  $\dMOT$ and $\bMOT$. In all three cases, for each $\bMOT$ value, $\Nfour$ shows a maximum (noted $N^*$) when varying $\dMOT$. The optimum is found at a negative detuning, $\dMOT^*$, whose magnitude $|\dMOT^*|$ increases with $\bMOT$. An overall optimum in $\Nfour$ is found at a finite value of the $(\bMOT,\dMOT)$ set.    

In the following, we compare the variations of $\dMOT^*$ and $N^*$ with $\bMOT$  quantitatively in the three aforementioned configurations in order to understand the capture process. We extract $\dMOT^*$ from the three sets of experimental data of Fig.\,\ref{Fig4} (a), (b), and (c) and display them as a function of $\bMOT $ in Fig.\,\ref{Fig4} (d). We observe that the variations of $\dMOT^*$ versus $\bMOT $ are similar in the presence of 421-nm and 626-nm push beams but slightly differ from the case without push beam. Furthermore, in all three cases, $\dMOT^*$ appears to obey a roughly linear dependence on $\bMOT$: $ \dMOT^* = \mu_R\bMOT +\delta_0$. Based on a simple theory of the 3D-MOT-capture process inspired from ref.\,\cite{Dalibard2015ubh} that we develop in Appendix~\ref{app:capture3D}  (see also Fig.\,\ref{Fig1} (d)), we can interpret the linear-dependence parameters in terms of effective capture parameters (see Eq.\,\eqref{Equ:3DMOTdvR}). The offset detuning relates to an effective capture velocity in the limit of small gradients ($b'\rightarrow 0$), $v_0$, via  $\delta_0=-\kred v_0/\sqrt{8}$, matching the relation expected in optical molasses. The slope relates to an effective capture radius with large gradients ($b'\rightarrow \infty$): $\mu_R \propto R_{\infty}$. A linear fit yields $v_0=[7.8(1), 7.5(1), 5.6(2)]\,$m/s and $R_{\infty}=[17(1), 19.1(7), 32(2)]\,$mm for [with 626-nm push, with 421-nm push, without push].  The increase in $v_0$ and decrease of  $R_{\infty}$  found when adding a push beam reveal a change in the velocity distribution of the atomic beam emerging from the 2D MOT. It evidences the boost in velocities and the decrease in spreading provided by the push beams.  

Our theory of the 3D-MOT capture process implies variations of the capture radius and velocity with $\bMOT$. In particular, $\vcap$ increases linearly with $\bMOT$. Ultimately, this limits the validity of our description to an intermediate gradient range as the capture parameters are bounded by the geometry and maximal radiative force, see Sec.\,\ref{sec:setup} and Appendix~\ref{app:capture3D}. These variations also enable us to comprehend the changes with $\bMOT$ in the loading efficiency  and therefore in $N^*$. Let us first describe the experimental observations. Figure\,\ref{Fig4} (e) depicts the variations of $N^*$ with $\bMOT$, as extracted from Figs.\,\ref{Fig4} (a), (b), and (c). We observe that $N^*$ varies with $\bMOT$ following a similar trend between the three configurations: Starting from small $\bMOT$, $N^*$ sharply increases and then slowly decreases when increasing $\bMOT$. 
A maximum of $N^*$ is found at an intermediate $\bMOT$ whose value depends on the push-beam configuration. The optimum $\bMOT$ is the lowest when no push beam is used and takes the value $\bMOT=0.31\,$G/cm. The optimum is shifted to larger values when using a push beam, namely $\bMOT =0.42$\,G/cm ($0.51$\,G/cm) with the 626-nm (421-nm) push beam. We also observe a steeper decrease of $N^*$ at large $\bMOT$ with a push beam compared to the case without push beam. 
Let us now understand this behavior based on the capture-process theory. The increase of $N^*$ at small $\bMOT$, is justified by the corresponding increase of $\vcap$, enabling to capture a larger fraction of the atomic distribution. The gain earned by increasing $\vcap$ saturates once it encompasses the full velocity distribution of the atomic beam or reaches the upper bound of $\vcap \lesssim 11\,$m/s imposed by our 3D-MOT configuration as introduced in Sec.\,\ref{sec:setup}. For larger values of $\bMOT$, a decrease of the capture efficiency is foreseen since the radiation pressure profile is not anymore optimized for the low velocities of the atomic jet. The weaker dependence of $N^*$ on $\bMOT$ in the absence of a push beam at large $\bMOT$ shall relate to the different velocity distributions in the atomic jet between the configurations.     
Using the relation of Appendix \ref{app:capture3D}, we can estimate the capture velocities for the optimal $\bMOT$ to $\vcap=[10.1(2), 10.7(2), 8.8(3)]$\,m/s for [626-nm push, 421-nm push, without push]. The different optimal $\bMOT$ can therefore be interpreted as a requirement to increase the capture velocities when introducing the push beams. 

Overall, the largest $N^*$ is found with the 626-nm push beam. The optimal values of the parameters determined within our optimization process are reported in Table \ref{tab:mot_opt_param}. Finally, we study the full loading curve of the 3D MOT in this identified optimal configuration. Additionally, we investigate the effect of changing the power of the 3D-MOT beams $\PMOT$. The loading curves and the extracted loading rates are displayed in Fig.\,\ref{Fig4} (f). At the previously set value of $\PMOT \approx 85$\,mW, we measure a loading rate of $\phiMOT = 1.10(2)\times 10^8$\,atoms/s and a saturation number of $\Nsat = 2.80(2) \times 10^8$. 
 By increasing $\PMOT$, both the saturation atom numbers and loading rates first sharply increase and then continue increasing at a slower rate. With the maximal power accessible in the present setup of $\approx 130\,$mW per beam, we find a maximal loading rate of $\phiMOT = 1.28(4)\times 10^8$ atoms/s and a saturation atom number of $\Nsat = 3.76(9) \times 10^8$. We note that a gain in the loading rate and saturation number seems still possible by increasing the power yet further.  In this work, we have simply relied on the power broadening of the 3D-MOT radiative force to increase the capture of our narrow-line 3D MOT. Additional schemes, such as spectral broadening~\cite{Vaidya2015dbf,Saskin2019nlc,Maier2014nlm,Dreon2017oca} or angled-slowing beams~\cite{Seo2020epo,Lunden2020etc}, could be considered for potential further performance enhancement. {We also note that the available power for the 2D-MOT cooling beam is presently technically limited, and we expect that a larger available power will enhance the present performance. The 3D-MOT loading achieved while reducing $\PtwoMOT$ is reported in App.~\ref{app:2DPow}. Finally, we stress that } increasing the oven temperature drastically increases the loading rate. As declared earlier, we decided to proceed with the optimization of our setup at a relatively low oven reservoir temperature for enhancing the lifetime of our source.



\begin{table}[h]

\centering

\begin{tabularx}{8.6cm}{lYYr}

\hline
\hline
     & Parameter &  Value & Unit\\
\hline
\multirow{5}{5em}{2D MOT} & $\lambda_{\rm 2D}$ & 421 & nm \\
& $\PtwoMOT$ & 430 & mW \\
& $w_{\rm 2D}$ & 16 & mm \\
& $\btwoMOT$ & 26.7 & G/cm \\ 
& $\dtwoMOT$ & -1.95 & \Gblue \\
& {$s_{\rm 2D}$} & {1.9} \\
\hline    
\multirow{4}{4em}{Push} & $\lambda_{\rm push}$ & 626 & nm \\
& $\Ppush$ & 18 & mW \\ 
& $w_{\rm push}$ & 0.8 & mm \\ 
& $\dpush$ & -82.3 & \Gred \\
& {$s_{\rm push}$} & {25000} \\
\hline
\multirow{5}{4em}{3D MOT (loading)} & $\lambda_{\rm 3D}$ & 626 & nm \\
& $\PMOT$ & 85 & mW \\
& $w_{\rm 3D}$ & 12 & mm\\
& $\bMOT$ & 0.42 & G/cm \\ 
& $\dMOT$ & -42.6 & \Gred \\ 
& {$s_{\rm 3D}$} & {520} \\
\hline 
\hline 
\end{tabularx}

\caption{Values of the different relevant parameters for optimal operation of our 2D-/3D-MOT scheme. The waists $w_{\rm 2D}$, $w_{\rm push}$, $w_{\rm 3D}$,  and the powers $\PtwoMOT$, $\PMOT$ were simply set to their values and not optimized. Other values are the results of our optimization process described in this paper.}
\label{tab:mot_opt_param}

\end{table}

\section{Conclusions} 
We have demonstrated the successful operation of a Dy intercombination line 3D MOT loaded from a 2D MOT working on the broad 421-nm transition. The addition of a push beam operating close to the intercombination line allows for the best loading performance, with a more-than-three-fold increase in the loading rate. We observe loading rates of $\phiMOT > 1\times 10^8$ atoms/s and a saturation number of $\Nsat \approx 3 \times 10^8$. This is similar or better compared to other cold-atom Dy setups based on Zeeman slowers despite the lower oven temperatures at which we operate, see e.g.~\cite{Youn2010dmo,Lu2010tud,Maier2014nlm,Dreon2017oca,Ilzhofer2018tsf,Lunden2020etc}. We note that the loading of the intercombination-line 3D MOT is a promising first step for quantum gas experiments. In particular, following the loading, a compression step can be applied in which the power and the detuning absolute value of the 3D-MOT beams are decreased~\cite{Maier2014nlm,Dreon2017oca,Ilzhofer2018tsf}. {By applying such a step to our samples, temperatures of $15\,\mu$K have been achieved with negligible atom loss.} 

We note that our setup has several advantages compared to Zeeman-slower-based ones. These include its compactness (our system is less than 1\,m long), its lower energetic consumption thanks to the use of permanent magnets and lower oven temperature, as well as the absence of a direct view between the oven and the center of the science chamber which reduces collisions with hot atoms without the need of a mechanical shutter and allows for a full optical switching of the atomic beam compatible with fast-cycling experiments. In future developments, a glass cell could directly be substituted for the metallic chamber and therefore allow for even greater optical access and faster magnetic-field control without e.g.\,the need for additional transport of the atomic cloud.

{Another interesting advantage of 3D MOTs of heavy atoms working on narrow transitions is the possibility to remove the MOT beam coming from the top, see e.g.~\cite{Ilzhofer2018tsf}. This setting allows for greater optical access, and e.g.\, easier integration of a high-resolution objective. In our setup, we observed that a 3D-MOT could also be loaded in such a 5-beam configuration but we did not perform an optimization in this setting yet.}


Moreover, we note that we achieved and observed 2D MOTs of other isotopes of Dy, namely $\Dyo$ and $\Dyf$. We also tried and successfully loaded a 3D MOT of $\Dyf$. $\Dyff$ could not be loaded on a first try and we suspect that a repumping frequency should be added to the 2D-MOT light. 

Our novel scheme thus constitutes a very favorable platform on which to build more complex experiments. To cite only two examples, based on such a 3D MOT, one could proceed with (i)  loading a dipole trap and performing evaporative cooling to quantum degeneracy, or (ii) loading single atoms in arrays of optical tweezers. Both platforms are highly promising candidates for quantum simulation or quantum computation purposes~\cite{Bloch2008mbp,Bloch2012qsw,Gross2017qsw,Kaufman2021quantum}. Finally, we also note that this scheme should be readily adaptable to other open-shell lanthanide species such as Er.

\vspace{5pt}
\vspace{2pt}

\textit{We note that another setup based on a similar 421-nm 2D-MOT loading a 626-nm 3D-MOT of Dy atoms has been developed in the group of I. Ferrier-Barbut (Bloch et al., in prep). We have greatly benefited from exchanges between our groups.}

\begin{acknowledgments}
First and foremost, we thank the Heidelberg quantum-gas community for their constant technical and scientific support along with the numerous fruitful discussions. This includes S.~Jochim, M.~Weidemüller, M.~Oberthaler, F.~Jendrzejewski, and their groups with a special mention to the HQA for sharing many of their design thoughts. We thank I.~Ferrier-Barbut and his group for open exchanges and discussions especially during the design process. We thank M. Barbiero for providing his simulation code. We thank J.~Wilson and J.~Thompson for enlightening discussions based on their Yb 2D/3D-MOT setup. We further thank A.~Patscheider, D.~Petter, G.~Natale, P.~Ilzhöfer, E.~Kirilov, J.~Beugnon, J.~Dalibard, and C.~Weitenberg for numerous early-stage discussions and technical advice. We thank T. Yefsah and M. Rabinovic for sharing technical designs. We thank L.~Hoenen, P.~Holzenkamp, V.~Salazar Silva, B.~Bader for their technical assistance. 

This work is funded by the European Research Council (ERC) under the European Union’s Horizon Europe research and innovation program under grant number 101040688 (project 2DDip), and by the Deutsche Forschungsgemeinschaft (DFG, German Research Foundation) through project-ID
273811115 (SFB1225 ISOQUANT) and under Germany’s Excellence Strategy EXC2181/1-390900948 (the Heidelberg Excellence Cluster STRUCTURES). Views and opinions expressed are however those of the authors only and do not necessarily reflect those of the European Union or the European Research Council. Neither the European Union nor the granting authority can be held responsible for them. J.\,G. acknowledges support from the International Max Planck Research School for Quantum Dynamics (IMPRS-QD). 
\end{acknowledgments}
\vspace{5pt}
$\dagger$ These authors contributed equally. 

$\star$ Correspondence and requests for materials should be addressed to chomaz@uni-heidelberg.de.

\appendix
\renewcommand\thefigure{\thesection S\arabic{figure}}   
\setcounter{figure}{0}   

{\section{3D-MOT loading as a function of the 2D-MOT power}\label{app:2DPow}
In the main text, we use the maximal power available of 421-nm light for our 2D-MOT cooling beam. This is done in order to maximize the loading performance while we expect the saturated regime not to be achieved for the 2D-MOT performances on the broad transition, see also~\cite{Tiecke2009hft,Lamporesi2013chf,Nosske2017tdm,Barbiero2020sec}. In this appendix, we report on the change in the loading performances with 2D-MOT cooling-beam power, $\PtwoMOT$. Fig.\,\ref{FigA} shows the relative variations in the loading rate $\phiMOT$ and saturation atom number $\Nsat$ with power together with their fits to exponential-growth function $A\left(1-e^{-(\PtwoMOT-P_0)/P_{\rm sat}}\right)$. The fit results show that the saturation regime is not achieved for the available power and increased performances are expected for increased power. 
\begin{figure}
    \includegraphics[width = 0.45\textwidth]{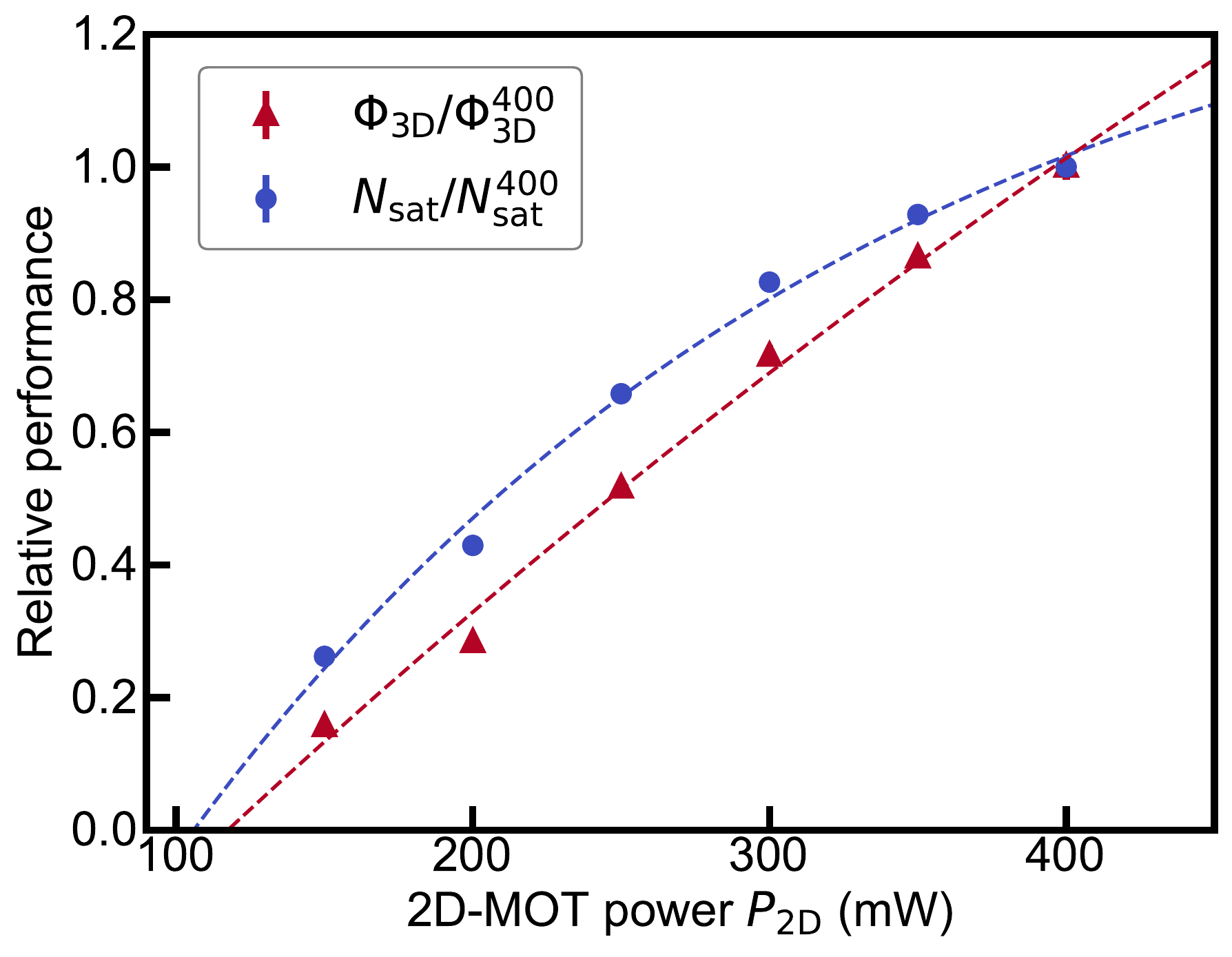}
    \caption{\label{FigA} \textbf{2D-MOT power dependence}. Relative values of the experimental 3D-MOT loading rate $\phiMOT$ and saturation atom numbers $\Nsat$ as a function of the 2D-MOT power $\PtwoMOT$. The values at $\PtwoMOT=400$mW are used as references.  The parameters are $\dpush=-46.54(13)\,\Gred$, $\Ppush=12.5$\,mW, and $\bMOT  = 0.57$\,G/cm, $\dMOT = -39.19(13)\Gred$ and $\phiMOT^{(400)}=8.1(2)\times 10^{7}$, $\Nsat^{(400)}=2.28(1)\times 10^{8}$. The errorbars are the $63\%$ confidence interval from the fit. The lines are exponential growth fits yielding $A= [3.84, 1.42]$ $P_0= [117, 106]$\,mW and $P_{\rm sat}= [923, 234]$\,mW, respectively.}
\end{figure}
}

\section{2D-MOT Simulation}\label{app:sim2D}

To estimate the performances of our 2D-MOT scheme as a function of the values of its parameters we performed Monte Carlo simulations of the trajectories of the atoms within the radiative-pressure force field, closely following the line of ref.~\cite{Barbiero2020sec}. Here we review the essential elements of the simulation, for more details see~\cite{Jianshun2022thesis}. 

The simulation starts at the oven output at time $t=0$. Here, the positions and velocities of $N_{\rm sim}$ (typically $N_{\rm sim}=50 000$) atoms are sampled assuming a uniform spatial distribution over the aperture and a thermal ($T=800\,^{\circ}$C) velocity distribution emerging from a tubed aperture. We use the transition regime approximation to describe its angular distribution~\cite{Giordmaine1960,Olander1970}. The distribution is cut to $7.5^{\circ}$ and only velocity $v<200\,$m/s are simulated. 

At times $t>0$, the atoms are submitted to the force field that results from the four passes of the 2D-MOT cooling beam via:
\begin{equation}\label{Equ:MOT force}
		\bs{F}_\text{2D}(\bs{r},\bs{v})= \sum_{i=1}^{4}\hbar \bs{k}_i \frac{\Gblue}{2} \frac{s_i(\bs{r},\bs{v})}{1 + \sum_{i} s_i(\bs{r},\bs{v})},
\end{equation}
where $\bs{k}_i$ is wavevector of the beam at pass $i$ ($\bs{k}_i= \kblue\frac{\pm\bs{e}_y \pm \bs{e}_z}{\sqrt{2}}$ with $\bs{e}_{x,y,z}$ the unit vector in direction $x,y,z$) and $s_i$ is given by
\begin{equation}\label{Equ:saturation parameter}
		s_i (\bs{r},\bs{v})= \frac{s_{i,0}(\bs{r})}{1+4\left(\delta_{\bs{r},\bs{v}}^{(i)}/\Gblue\right)^2},
	\end{equation}
with $s_{i,0}(\bs{r})= I_i(\bs{r})/I_{\rm sat}^{421}$ the local saturation parameter, {$I_i(\bs{r})=2 \PtwoMOT / \pi w_{\rm 2D}^2 \cdot \exp{(-2r^2/w_{\rm 2D}^2)}$} the local intensity~\cite{Wohlleben2001}.   
The local detuning $\delta_{\bs{r},\bs{v}}^{(i)}$ in beam-pass $i$ is position- and velocity-dependent due to the Zeeman effect in the 2D-MOT magnetic field $\bs{B}_{\rm 2D}(\bs{r})$ and the Doppler effect:   
\begin{equation}\label{Equ:MOT detuning}
		\delta_{\bs{r},\bs{v}}^{(i)} = \dtwoMOT - \bs{k}_i \cdot \bs{v} + \frac{\mu_B}{\hbar} g_{J'}^{421}\Delta m_i |\bs{B}(\bs{r})|,
	\end{equation}
where $\Delta m_i=\pm 1$ depending if the beam pass $i$ drives the $\sigma_\pm$ transition (i.e. defined by the sign of $\bs{B}\cdot \bs{k}_i$). We use an ideal magnetic environment with $\bs{B}_{\rm 2D}(\bs{r})=\btwoMOT(y\bs{e}_z+z\bs{e}_y)$.

The position $\bs{r}$ and velocity $\bs{v}$ of each atom are computed in an iterative manner using a four-step Runge-Kutta method with a time step $\Delta t = 50$\,$\rm{\mu s}$~\cite{scherer2010}. At each time step, Newton's second law is applied with Eq.\,\eqref{Equ:MOT force} providing the instantaneous acceleration. Besides Eq.\,\eqref{Equ:MOT force}, we also include the effect of spontaneous emission approximately by an additional random force over the time step $\Delta t$:
\begin{equation}\label{Equ:spont force}
		\bs{F}_\text{spont}(\bs{r},\bs{v})= \hbar \kblue \sqrt{ \frac{\Gblue}{2} \frac{\sum_i s_i(\bs{r},\bs{v})}{(1 + \sum_{i} s_i(\bs{r},\bs{v}))} \Delta t }\hat{\bs{e}}_\text{spont},
\end{equation}
where $\hat{\bs{e}}_\text{spont}$ is a random unit vector, which we randomly draw at each time step of the simulation~\cite{Lett1989, Kohel2003}. 
We note that Eqs. (\ref{Equ:MOT force}-\ref{Equ:spont force}) approximate the complex  $J=8 \rightarrow J'=9$ structure of Dy to a simpler $J=0 \rightarrow J'=1$ structure. This accounts for the light polarization effect but neglects the spinful character of the ground state. Furthermore, the non-closed character of the cooling transition is omitted~\cite{Lu2010tud}. 

We stop the simulation after a total time of $10$\,ms and count the number of atoms $N_{\rm cap}$ whose trajectories are within $25\,$mrad solid angle from the $x>0$-axis. We then extract the 2D-MOT flux via $\phitwoMOT=\phi_{\rm sim}(T) N_{\rm cap}/N_{\rm sim}$ where $\phi_{\rm sim}(T)$ is the flux of atoms exiting the oven at the temperature $T$ accounting for the limits set in angle and magnitude for the velocities of the simulated atoms. For $T=800\,^{\circ}$C ($T=1000\,^{\circ}$C), $\phi_{\rm sim}(T)=1.5\times 10^{10}$\,atoms/s ($\phi_{\rm sim}(T)=1.9\times 10^{12}$\,atoms/s) for $^{164}$Dy. The actual oven temperature certainly lies in between these two values due to the higher hot lip temperature also partly heating the reservoir. The maximum flux $\phitwoMOT$ found in Fig.\,\ref{Fig2}, using $T=800\,^{\circ}$C, is estimated in absolute value to $\phitwoMOT= 3\times 10^{8}$\,$^{164}$Dy atoms/s while using $T=1000\,^{\circ}$C, we find a maximum $\phitwoMOT= 3.5\times 10^{10}$\,$^{164}$Dy atoms/s. Therefore the fraction of loaded trajectories in our simulation is on the order of $2\%$ and very slightly larger at lower oven temperatures. We also note that the total oven flux is about three orders of magnitude larger than $\phi_{\rm sim}$.

Simulations further including the push beam and 3D-MOT beams and counting the number of atoms loaded in the 3D MOT were also performed. In this case, the force \eqref{Equ:push_force} is added to the 2D-MOT force, and another force similar to \eqref{Equ:MOT force} but summing over the six 3D-MOT beam passes and centered at $x_0=+347$\,mm is included. In this case, we count the number of atoms trapped within a 10\,mm radius around $x_0$ while extending the simulation time to $260$\,ms. We then calculate the simulated $\phiMOT$ similarly to $\phitwoMOT$. Using a push beam we find that the fraction of loaded trajectories is again reduced by a factor of $\sim 20$ to $30$ compared to the 2D-MOT results, yielding very weak signals, $N_{\rm cap}$ of the order of ten with the used $N_{\rm sim} = 50000$. This weak signal precludes reliable simulation results for the 3D-MOT loading as highlighted in Secs.~\ref{sec:setup} and~\ref{sec:optim2DMOT}.

\section{Basic theory of the 3D-MOT capture process}\label{app:capture3D}

We consider the process in which a 3D MOT with a 6-beam geometry similar to the sketch of Fig.\,\ref{Fig1} (d) captures atoms from an atomic beam propagating along $+x$ axis. A simple understanding can be drawn, following the lines of ref.~\cite{Dalibard2015ubh}, based on the phase-space-dependent detuning of the cooling beams. Accounting for our beam geometry, the detuning of an atom of position $x'$ and velocity $v_x$ writes:
\begin{equation}\label{Equ:3DMOT detuning}
		\delta_{x',v_x}^{\pm} = \dMOT \pm \left(\frac{\kred v_x}{\sqrt{2}}+\frac{\mu_{626}'\bMOT x'}{\hbar}\right),
	\end{equation}
 where $+$ ($-$) holds for the two cooling beams propagating against (along) the atomic beam. Here, $x'$ denotes the position of the atoms with respect to the science chamber center ($x'=x-x0$), $\mu'_{626}=g_{J'}^{626}\mu_B$ is the 626nm-transition excited state differential magnetic moment and $\mu_B$ the Bohr magneton. The capturing process mostly relies on the effect of the two beams propagating against the atomic beam since it can be seen from Eq.\,\eqref{Equ:3DMOT detuning} that they are brought closer to resonance for $v_x>0$ (i.e.\,as in the atomic beam), or for $x'>0$ (i.e.\,for atoms that would go beyond the trap center). Hereafter, our capturing model neglects the effect of the two other beams and simply uses $\delta_{x',v_x}^{+}$.  

We define $\vcap$ as the maximal velocity of the atoms that can be stopped within the capture region of radius $\Rcap$. Here, we assume that both $\vcap$ and $\Rcap$ depend on $\bMOT$.  Following ref.~\cite{Dalibard2015ubh}, we then formulate that the optimal capture configuration is achieved by adjusting $\dMOT$ and $\bMOT$ such that the atoms with a velocity $v_x= \vcap (\bMOT)$ are on resonance at the entrance and exit of the capture region (see highlighted points in Fig.\,\ref{Fig1} (d)). This yields $\delta_{x'=-\Rcap,v_x=\vcap (\bMOT)}^{+}=0$ at the entrance, and $\delta_{x'=\Rcap,v_x=0}^{+}=0$ at the exit. This ensures a maximal radiative force for this class of atoms. The entrance and exit conditions respectively yield:
\begin{eqnarray}
  \dMOT^*(\bMOT )&\approx& \frac{\mu_{626}'\bMOT \Rcap(\bMOT )}{\hbar}-\frac{\kred \vcap (\bMOT )}{\sqrt{2}},\\
  \dMOT^*(\bMOT )&\approx& -\frac{\mu_{626}'\bMOT \Rcap(\bMOT )}{\hbar}.
	\end{eqnarray}
where $\dMOT^*(\bMOT )$ is the value of the detuning providing the optimal capture configuration for the gradient $\bMOT$.  Combining these relations yields 
\begin{equation}\label{Equ:3DMOTdvR}
		\dMOT^*(\bMOT )\approx -\frac{\mu_{626}'\bMOT \Rcap}{\hbar}\approx-\frac{\kred \vcap}{\sqrt{8}}.
	\end{equation}
The approximate relation \eqref{Equ:3DMOTdvR} can be combined with the linear dependence observed experimentally, $\dMOT^*(\bMOT)= \mu_R\bMOT +\delta_0$, and allows giving a simple interpretation of the parameters $\mu_R$ and $\delta_0$. The offset value at $\bMOT =0$, $\delta_0= \dMOT^*(\bMOT =0 )$ can be written as $\delta_0=-\kred v_0/\sqrt{8}$ with $v_0$ the effective capture velocity at small gradients $\bMOT\rightarrow 0$, matching the formula expected in optical molasses~\cite{Dalibard2015ubh}. The scaling factor $\mu_R$ can instead be written as $\mu_R =-\mu_{626}'R_{\infty}/\hbar$ with $R_{\infty}$ the effective capture radius of the 3D-MOT with large gradient, $\bMOT \rightarrow \infty$. 

The linear relation can also be combined with \eqref{Equ:3DMOTdvR} to extract the dependence of the capture velocity and radius on $\bMOT$: 
\begin{eqnarray}
\label{Equ:vcap}
  \vcap (\bMOT )&=&v_0 + \frac{\sqrt{8}\mu_{626}'\bMOT R_{\infty}}{\hbar \kred},\\
   \label{Equ:Rcap}
   \Rcap(\bMOT )&=&R_{\infty}+\frac{\hbar \kred v_0}{\sqrt{8}\mu_{626}'\bMOT}.
\end{eqnarray}
We find that $\vcap$ increases and $\Rcap$ decreases with the MOT gradient, which matches intuitively expected behaviors. We note that both relations present nonphysical divergences, either at large gradients for $\vcap$ or small gradients for $\Rcap$, which evidence the limitations of our simple model. More precisely, the value of $\Rcap$ and $\vcap$ are ultimately bounded by our beam geometry and the maximal force that can be applied, as discussed in Sec.\,\ref{sec:setup}. 

\end{document}